\documentclass[aip,amsmath,amssymb,reprint]{revtex4-1}

\usepackage[dvipsnames]{xcolor}
\usepackage{graphicx}
\usepackage{dcolumn}
\usepackage{bm}
\usepackage{dcolumn}
\usepackage{epstopdf}
\usepackage{xcolor}
\usepackage{natbib}
\usepackage{subcaption} 
\usepackage{float}
\usepackage{tabularx}
\usepackage{mathrsfs}
\usepackage{upgreek}
\usepackage[a]{esvect}
\setcitestyle{square}
\usepackage[toc,page]{appendix}
\usepackage[font={small,normalfont}]{caption}
\usepackage[colorlinks=true, allcolors=blue]{hyperref}
\usepackage[normalem]{ulem}

\usepackage{upgreek}
\usepackage[capitalise]{cleveref}
\newcommand{\micron}{{\upmu\mathrm{m}}}

\begin{document}

\title{Effects of simulation dimensionality on laser-driven electron acceleration and photon emission in hollow micro-channel targets}

\author{Tao Wang}
\affiliation{Department of Mechanical and Aerospace Engineering, University of California at San Diego, La Jolla, CA 92093}

\affiliation{Center for Energy Research, University of California at San Diego, La Jolla, CA 92093}

\author{David Blackman}
\affiliation{Department of Mechanical and Aerospace Engineering, University of California at San Diego, La Jolla, CA 92093}

\affiliation{Center for Energy Research, University of California at San Diego, La Jolla, CA 92093}

\author{Katherine Chin}
\affiliation{Department of Mechanical and Aerospace Engineering, University of California at San Diego, La Jolla, CA 92093}

\affiliation{Center for Energy Research, University of California at San Diego, La Jolla, CA 92093}

\author{Alexey Arefiev}
\affiliation{Department of Mechanical and Aerospace Engineering, University of California at San Diego, La Jolla, CA 92093}

\affiliation{Center for Energy Research, University of California at San Diego, La Jolla, CA 92093}

\date{\today}

\begin{abstract}
Using two-dimensional (2D) and three-dimensional (3D) kinetic simulations, we examine the impact of simulation dimensionality on the laser-driven electron acceleration and the emission of collimated $\gamma$-ray beams from hollow micro-channel targets. We demonstrate that the dimensionality of the simulations considerably influences the results of electron acceleration and photon generation
owing to the variation of laser phase velocity in different geometries. In a 3D simulation with a cylindrical geometry, the acceleration process of electrons terminates early due to the higher phase velocity of the propagating laser fields; in contrast, 2D simulations with planar geometry tend to have prolonged electron acceleration and thus produce much more energetic electrons. The photon beam generated in the 3D setup is found to be more diverged accompanied with a lower conversion efficiency. Our work 
concludes that the 2D simulation can qualitatively reproduce the features in 3D simulation, but for quantitative evaluations and reliable predictions to facilitate experiment designs, 3D modelling is strongly recommended. 
\end{abstract}

\maketitle


\section{Introduction} \label{Sec-1}

\begin{figure*}[!htp]
\centering
    \includegraphics[width=0.9\textwidth,trim={1cm 0.4cm 0cm 0cm},clip]{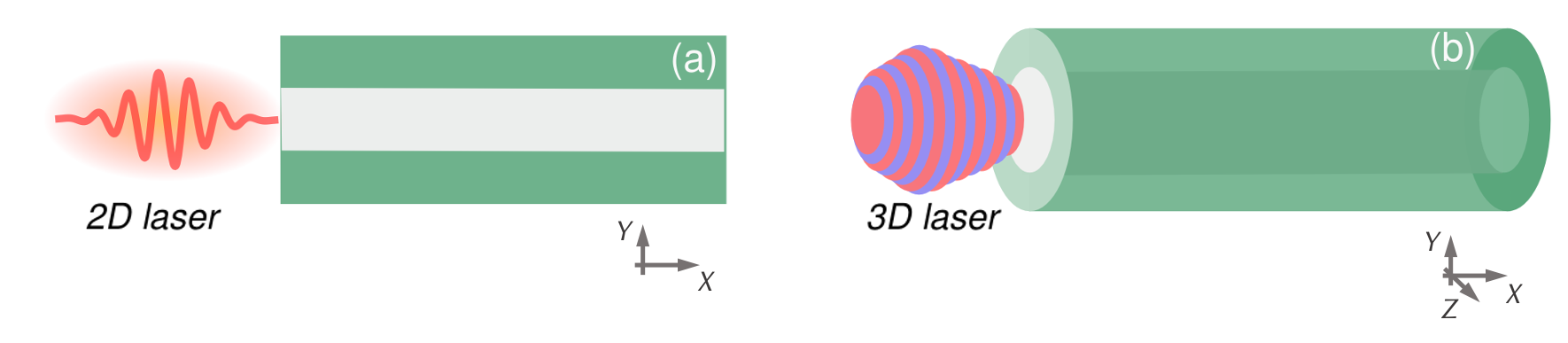}
    \caption{\label{fig:Schematics} Laser-irradiated hollow targets in 2D and 3D. (a) 2D setup where a linearly polarized two-dimensional Gaussian pulse interacts with a two-dimensional hollow plasma channel. (b) 3D setup where a linearly polarized and cylindrically symmetric Gaussian pulse interacts with a hollow cylindrical target.}
\end{figure*}

The interaction of high-power ultra-intense lasers and structured (both nanostructured~\cite{nano_2004,Nishikawa_applied_nano2004,nano_2011_pop,nanosphere_2012_PRL,Blanco_2017_nanostructured,Moreau_2019_nanowire,nanowire_2021_SR} and microstructured~\cite{Attosecond_Naumova,Klimo_2011_micro,Xiao_hollow_channel,microwire_hedp,Feng_micro_effects,Jiang_microengineering,ji2016towards,Stark2016PRL,yu2018generation,Snyder_microchannels,microtube_Beg_PRE}) targets has been a topic of great interest for its capability of enhancing the laser energy conversion efficiency~\cite{nano_2004}, high-order harmonics generation~\cite{HHG_pop_2010,HHG_2018_PRL}, charged particles (relativistic electrons and ions~\cite{nano_2011_pop,nanosphere_2012_PRL,microtube_Beg_PRE}) acceleration and the production of X-ray~\cite{nano_2004,Xray_micro_2016_PRL,xray_2017_pop,Rolles_xray_nano_2018} to $\gamma$-ray~\cite{Stark2016PRL,yu2018generation,gammaray_2018_pnas} radiation. The produced charged particle and photon beams have a wide range of applications from medical ion therapy~\cite{malka2004practicability,robson2007scaling}, nuclear physics~\cite{bychenkov1999nuclear,Bremsstrahlung_NRF} to photon-photon pair production~\cite{yu_2019,wang_pair_comment,PRApplied_power}. The microstructured targets with characteristic size of surface modulation comparable to laser wavelength are able to extensively absorb laser energy through various processes, including surface plasmon resonance excitation~\cite{PRL_nano_plasmon_2003}, multipass stochastic heating~\cite{Breizman_Alex_pop_2005} in dense clusters, prolonged acceleration distance in hollow channels~\cite{Xiao_hollow_channel} and microwires~\cite{microwire_hedp} and relativistic transparency in prefilled channels~\cite{Stark2016PRL}. In this paper, we examine the regime involving hollow micro-channels (see \cref{fig:Schematics}).  

When ultra-intense lasers irradiate hollow micro-channels, strong laser fields directly act on electrons, dragging them into the channel and forming periodic electron bunches which then surf along with the laser pulse, gaining energy from the laser. Additionally, the presence of a channel guides the propagation of the electromagnetic fields, confines the electron motion and as a result, leads to a well collimated photon emission. This setup can serve as a promising electron source to further stimulate ion acceleration~\cite{microtube_Beg_PRE} as long as the ion expansion does not significantly impact electron acceleration~\cite{Wang_2019}. However, to successfully apply such an electron source in experiments, careful numerical investigations are needed in order to determine where the electron energy peaks, i.e. the location to cut off the channel and collect an electron source with an optimal spectrum. 

To carry out such numerical studies, one can choose between 2D3V and 3D3V Particle-In-Cell (PIC) simulations. Both 2D~\cite{Klimo_2011_micro,Xiao_hollow_channel,Attosecond_Naumova,zou2017microchannel,Gong_hollow_channel} and 3D~\cite{ji2016towards,Jiang_microengineering,Jiang_PRE,Snyder_microchannels,Serebryakov_2019_3Dstructured} numerical simulations have been widely used to characterize laser interactions with structured targets. The appeal of 2D simulations is that they require significantly less computational resources than 3D simulations, so one is able to perform extensive parameter scans using 2D simulations. However, the field topology differs between 2D and 3D setups and it is not immediately clear how the differences impact the particle dynamics. It is then important to evaluate the dimensionality effects on a case-by-case basis. A few publications~\cite{stark_dimensionality_ion,Liu_dimensionality_ion,Xiao_2018_dimensionality_ion,Blanco_2017_nanostructured} have discussed the influences of simulation dimensionality on ion acceleration with various target geometries. But, to our knowledge, nobody has examined and explained the physics of dimensional effects on laser-irradiated hollow micro-channel targets.

In this paper, we show that the chosen dimensionality has a considerable effect on electron acceleration and the associated photon emission. First, we provide theoretical analysis to demonstrate that the dephasing rate between the accelerated electron and laser wave-fonts strongly depends on simulation geometry. Later we show numerical evidence to demonstrate that the dephasing rate differs with simulation dimensionality and such a difference is the key reason for the distinguished observation in terms of electron and photon beam generation. Through collectively evaluating generated particles and detailed particle tracking, we show that the occurrence of high phase velocity in the 3D setup terminates electron acceleration process early in space and time, and leads to a reduction of photon emission. 

\section{Phase velocity in 2D and 3D waveguides}

Previous studies\cite{Xiao_hollow_channel,Wang_2019,Gong_hollow_channel} have shown that, in a hollow channel without significant ion expansion, the dominant contributor to electron acceleration is the electric field in the direction of the laser propagation, i.e. the longitudinal field $E_x$. The corresponding structure of propagating electromagnetic fields can be characterized as TM (transverse magnetic) modes. The wave equation of TM modes can be written as
\begin{equation}
    \bigg(\dfrac{\partial^2}{\partial y^2} + \dfrac{\partial^2}{\partial z^2}\bigg)E_x + \bigg(\dfrac{\omega^2}{c^2}-k^2\bigg)E_x = 0 ,
    \label{Eq:wave_equation}
\end{equation}
where $\omega$ is the wave frequency, $k$ is the wavenumber and $c$ is the speed of light. For a two-dimensional waveguide (shown in \cref{fig:Schematics}a), $E_x$ is a function of $y$ and $\partial E_x/ \partial z = 0$. \Cref{Eq:wave_equation} then becomes ${\partial^2}E_x/{\partial y^2} + (\omega^2/c^2-k^2) E_x =0$. We choose $E_x = E_0 sin(\pi y / R)$ as the TM-mode solution that matches the field structure in the incoming pulse (i.e. $E_x=0$ on axis) and the boundary conditions which requires $E_x(y = \pm R) = 0$. Here $R$ represents the radius of the plasma channel. The dispersion equation in the 2D waveguide is then given by

\begin{equation}
     \dfrac{\omega^2}{c^2} = k^2 + \dfrac{\pi^2}{R^2} .
     \label{Eq:2D_dispersion}
\end{equation}

For the 3D cylindrical waveguide (shown in \cref{fig:Schematics}b), it is convenient to rewrite \cref{Eq:wave_equation} in cylindrical form as 
\begin{equation}
    \dfrac{1}{r}\dfrac{\partial}{\partial r} \bigg(r \dfrac{\partial E_x}{\partial r}\bigg)
 + \dfrac{1}{r^2}\dfrac{\partial^2 E_x}{\partial \psi^2} + \bigg(\dfrac{\omega^2}{c^2}-k^2\bigg)E_x = 0,
 \end{equation}
where $r$ is the axial distance and $\psi$ is the azimuth. Assuming a solution in the form of $E_x = f \sin(\psi)$, the resulting equation for $f$ then reads 
\begin{equation}
    s^2 \dfrac{\partial^2 f}{\partial s^2} + s \dfrac{\partial f}{\partial s} + (s^2 -1) = f.
\end{equation} 
Here $s \equiv \beta r$ and $\beta^2 = \omega^2/c^2 -k^2$. The solution for $f$ is given by $f = E_{\parallel} J_1(s) = E_{\parallel}J_1(\beta r)$ where $J_1(x)$ is a Bessel function of the first kind and $E_{\parallel}$ is the amplitude of the longitudinal field. At the boundary of the cylindrical waveguide, $f(r=R) = 0$ which yields $\beta R \approx 3.8$. The dispersion relation of the 3D cylindrical waveguide
is then written as 
\begin{equation}
   \dfrac{\omega^2}{c^2} = k^2 + \dfrac{14.7}{R^2}.
   \label{Eq:3D_dispersion}
\end{equation}

It is convenient to derive a general expression for $v_{ph}$ of a propagating wave with the dispersion relations given in \cref{Eq:2D_dispersion,Eq:3D_dispersion}.
\begin{equation}
    u = \dfrac{v_{ph}}{c} = \dfrac{\omega}{kc} = \sqrt{1+\dfrac{\alpha^2}{R^2k^2}} \approx \sqrt{1+\dfrac{\alpha^2}{R^2k_0^2}},
    \label{Eq:combined_dispersion}
\end{equation}
where we set $k \approx k_0=\omega/c$, $\alpha^2 = \pi^2 \approx 9.9$ for the 2D channel and $\alpha^2 = 14.7$ for the 3D cylindrical channel. Subtracting the $v_{ph}$ from $c$, in the limit of $\alpha^2 \ll R^2 k_0^2$ we have
\begin{equation}
    \delta u = u - 1 \approx \dfrac{1}{2}\dfrac{\alpha^2}{R^2k_0^2}.
    \label{Eq:dephasing}
\end{equation}
$\delta u$ is a dimensionless parameter used to quantify the degree of supuerluminosity~\cite{Robinson_PoP_2015}; it can be understood as a dephasing rate, since it illustrates how quickly the local laser wave-front outpaces the electron in question.
The ratio of the dephasing rate for the case in 2D to that in 3D is
\begin{equation}
    \dfrac{\delta u_{3D}}{\delta u_{2D}} \approx 1.5.
\end{equation}
In the following sections we are going to demonstrate via numerical simulations that the lower dephasing rate in 2D leads to an overestimate of electron acceleration and $\gamma$-ray emission.

\section{Impact of dimensionality on electron acceleration}\label{impact_electron}

\begin{figure*}[htp!]
\centering
    \includegraphics[width=0.9\textwidth,trim={0cm 0.cm 0cm 0.0cm},clip]{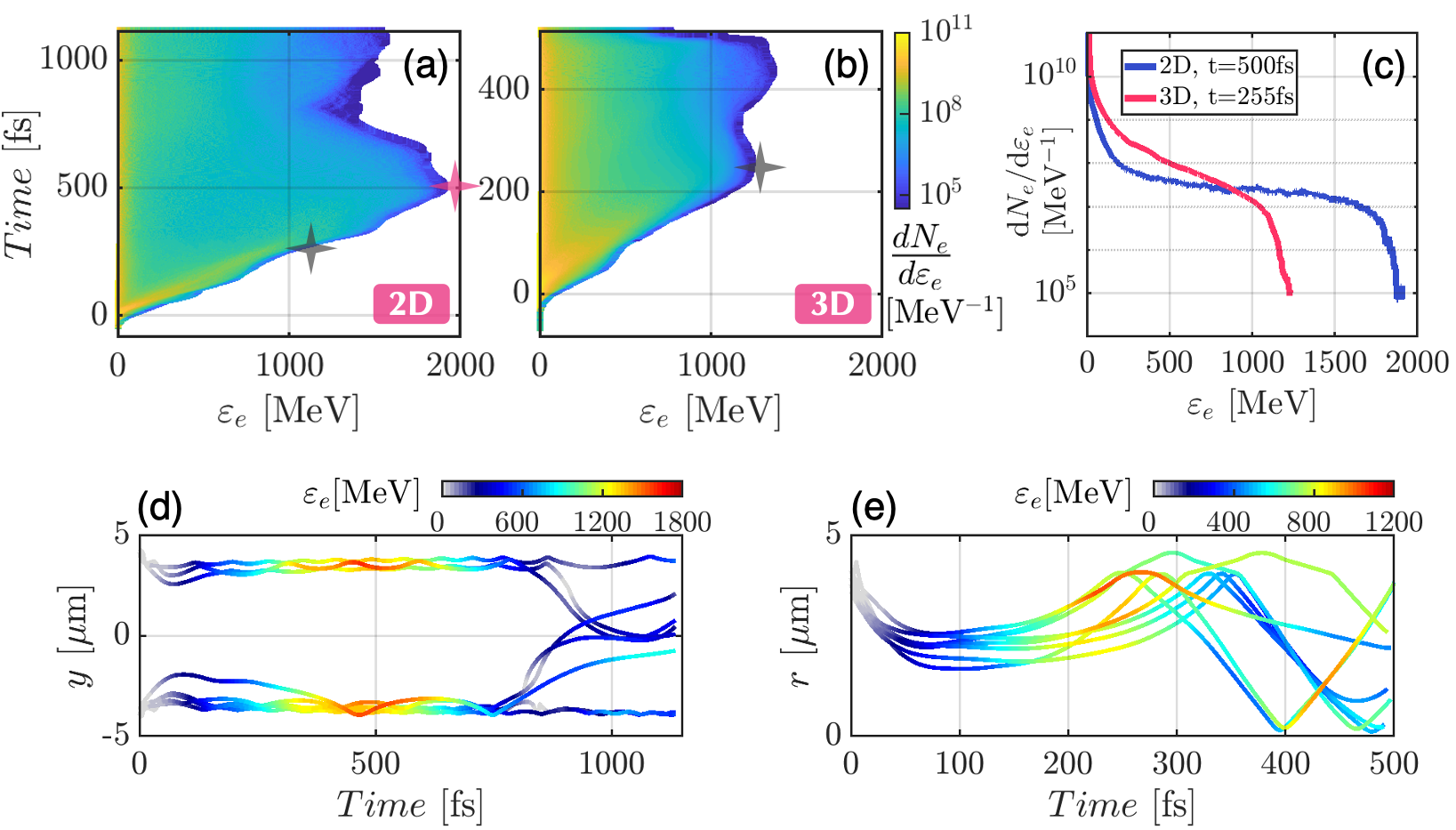}
    \caption{\label{fig:Electron_spectrum_2D3D} Time-resolved electron energy spectrum observed in (a) the 2D simulation (b) the 3D simulation. The colorbar represents the number of electrons collected onto a certain energy bin
    d$\varepsilon_e$. (c) The peak energy spectrum in 2D and 3D which occurred at $t=500$ fs (marked with red star) and $t = 255$ fs (marked with gray star) respectively. We define $t = 0$ fs as the time when the laser pulse reaches its peak amplitude in the focal plane at $x = 0$ $\mu$m in the absence of the target. Typical electron trajectories are plotted in (d) $y$-$t$ space for the 2D simulation (e) $r$-$t$ space for the 3D simulation. Here $r=\sqrt{y^2+z^2}$. The color on the trajectories represents electron relativistic energy.}
\end{figure*}

We model the electron acceleration through fully relativistic PIC simulations using the EPOCH code\cite{Epoch} which is available in both 2D and 3D. The target geometries implemented in the simulations are depicted in \cref{fig:Schematics}. In 2D, the target is a straight empty channel enclosed by two uniform plasma slabs. In 3D, the channel is enclosed by a cylindrical wall of plasma, with a channel diameter such that a slice along the target axis would be identical to the 2D setup. The 2D and 3D simulations share the same plasma composition, plasma density, laser intensity, temporal profile, and laser spot size. The laser intensity is set as $1.37 \times 10^{22} \ \rm{W/cm}^2$, corresponding to $a_0 = 100$. Here $a_0 \equiv |e|E_0/(m_e c\omega)$ is the normalized laser amplitude, where $E_0$ is the peak amplitude of the electric field in the incoming laser pulse, and $m_e$ and $e$ are the electron mass and charge. The laser pulse is always focused at the channel entrance. We choose gold as the original target material with a density of 1.5 g/cm$^3$. According to the field ionization model, the considered laser pulse is capable of ionizing gold atoms to the level of $Z = 69$. In the simulations, the target is pre-ionized accordingly to a plasma composed of Au$^{+69}$ (density 4 $n_{cr}$) and $e^-$ (density 276 $n_{cr}$) where $n_{cr} \equiv m_e\omega/(4\pi e^2)$ is the critical density. A detailed comparison of parameters used in the 2D and 3D simulations is listed \cref{table_PIC}. Note that although the target lengths are set differently, this is done such that in different dimensionalities the target is long enough for the electrons to reach their first energy peak. 

\begin{table}
\caption{Parameters used in 2D and 3D PIC simulations. $^*$To calculate the laser power, we assume the length of third dimension in 2D simulations as 2 $\micron$.}
\label{table_PIC}
\begin{tabular}{ |l|l|}
  \hline
  \multicolumn{2}{|c|}{Parameters shared by 2D and 3D simulations} \\
  \hline
  \multicolumn{2}{|l|}{\underline{\bf Laser pulse:} }\\
  Peak intensity & $1.37 \times 10^{22}$ W/cm$^2$ \\
  $a_0$ & 100\\
  Polarization & linearly along $\mathbf{\hat{y}}$\\
  Wavelength & $\lambda_L = 1~\micron$ \\
  Location of the focal plane & $x = 0~\micron$, surface of plasma\\
  Pulse temporal profile & Gaussian \\
  Pulse duration & \\
  (FHWM for intensity) & 30 fs\\
  Pulse width/focal spot  &  \\
  (FWHM for intensity) & $w_0 = 2.8~\micron$\\
  \multicolumn{2}{|l|}{\underline{\bf Plasma:} }\\
  Composition & gold ions and electrons \\
  Channel radius & $R = 4.0~\micron$ \\
  Target thickness & $d = 0.4~\micron$\\
  Electron density & $n_e = 276 n_{cr}$ \\
  Ion mass to charge ratio & $197 m_p$ : 69 \\
  Ion mobility & mobile \\
  \hline
  \multicolumn{2}{|c|}{Parameters varying in 2D and 3D simulations} \\
  \hline
  Spatial resolution & {{2D:}} $100/\micron \times 100/\micron$ \\
  & {{3D:}} $50/\micron \times 50/\micron \times 50/\micron$\\
  $\#$ of macro-particles/cell  &  {{2D:}} 100 for $e^-$, 5 for  $\rm{Au}^{+69}$\\
  & {{3D:}} 10 for $e^{-}$, 5 for $\rm{Au}^{+69}$\\
  laser geometry & 2D: symmetric about $y$-axis \\
  & 3D: cylindrical symmetry  \\
  laser power & 2D:  0.82 PW$^*$\\
  & 3D: 1.24 PW\\
  Target length & {{2D:}} $L = 350~\micron$ \\
  & {{3D:}} $L = 150 \micron$ \\
  \hline
\end{tabular}
\end{table}

\Cref{fig:Electron_spectrum_2D3D}(a) and (b) illustrate the time history of the electron energy spectrum observed in both 2D and 3D simulations. There exists more than one energy peak in both \cref{fig:Electron_spectrum_2D3D}(a) and \cref{fig:Electron_spectrum_2D3D}(b) and our focus is on the first energy peak which happens before any deceleration takes effect. It is clear that the maximum electron energy gain achieved in 2D significantly exceeds the gain observed in the 3D case. In fact, the first energy peak in 2D occurs at $t=500$ fs with $\varepsilon_e = 1920$ MeV while in 3D the peak occurs at $t=255$ fs with $\varepsilon_e =1240$ MeV. \Cref{fig:Electron_spectrum_2D3D}(c) gives a direct comparison of the peak spectra observed in the 2D and 3D simulations. In order to facilitate a comparison between the two simulations, we added a gray star to \cref{fig:Electron_spectrum_2D3D}(a) that represents the energy and time of the first peak from \cref{fig:Electron_spectrum_2D3D}(b). Evidently, the acceleration in 2D lasts longer, which results in a more energetic electron spectrum. Note that to make a quantitative comparison with the 3D simulations, we assume a uniform third dimension with a length of $2~\micron$ for the 2D simulation. 

To further understand the electron acceleration process, we tracked energetic electrons in both simulations. \Cref{fig:Electron_spectrum_2D3D}(d) and \Cref{fig:Electron_spectrum_2D3D}(e) illustrate the trajectories of representative electrons selected from both the 2D and 3D simulations. Plotted in the space of the transverse location ($y$ or $r$) and time, the trajectories are in agreement with the time history of electron spectrum given in \cref{fig:Electron_spectrum_2D3D}(a-b); the electron energies peak at the corresponding moments. Regardless of the dimensionality, the electrons surf along the channel wall while getting accelerated by longitudinal electric fields until reaching their first energy peak. However, the horizontal surfing of electrons in the 3D simulation terminates earlier due to its higher dephasing rate, which will be elaborated in the next section. It is worth noting that the second energy peak observed in \cref{fig:Electron_spectrum_2D3D}(a-b) is correlated with the electron motion of crossing the central axis, indicating an involvement of the transverse electric field in electron acceleration. 


Through evaluating the electron energy spectrum and tracking individual electron motion, we have shown that the dimensionality of simulations significantly impacts electron acceleration. The 2D simulations tend to extend the electron acceleration process, leading to an overestimate of the maximum electron energy when compared to more realistic 3D simulations.  

\section{Electric field profiles and phase velocity}

An electron travelling in the laser fields gains energy only while staying in the accelerating phase of the electric field. The electron can gain energy from both longitudinal ($E_{\parallel} = E_x$) and transverse ($E_{\perp} = E_y,\ E_z$) electric fields. The total work done by the electric fields on a given electron can be expressed as 
\begin{equation}
 W_{tot} = W_{\parallel} + W_{\perp} =  -|e| \int_{-\infty}^{t} (E_{\parallel}v_{\parallel} + E_{\perp}v_{\perp}) \,dt{'} .
\end{equation}
Figure 3 shows the contributions of $W_{\parallel}$ and $W_{tot}$ to electron relativistic energy at the first peak by binning all electrons according to their energies. In the 2D simulation 95\% of the energy of energetic electrons ($\varepsilon_e > 500$ MeV) comes from the work done by the longitudinal electric fields and in the 3D case the quantity is 86\%. It is then reasonable to approximate
\begin{equation}
 W_{tot} \approx W_{\parallel} =  -|e| \int_{-\infty}^{t} E_{\parallel}v_{\parallel} \,dt{'} .
\end{equation} 
This simply allows us to narrow down the investigation of electron acceleration to a single component of the electric fields.

\begin{figure} [htp!]
    \includegraphics[width=0.98\columnwidth,trim={0.2cm 0.1cm 0cm 0cm},clip]{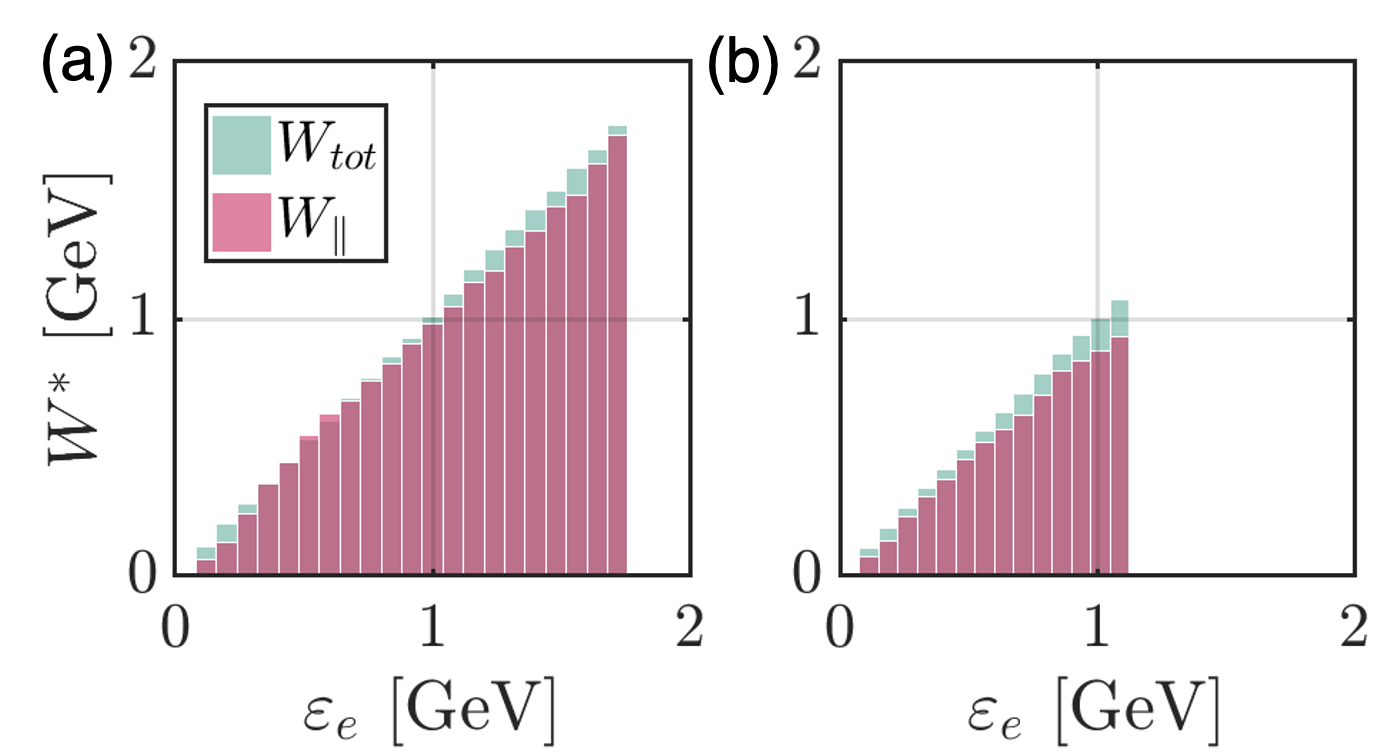}
    \caption{\label{fig:work_para_perp} Contributions to the electron relativistic energy by $E_{\parallel}$ ($W_{\parallel}$, red bars) and the electric fields (including both $E_{\parallel}$ and $E_{\perp}$, $W_{tot}$, green bars) in (a) the 2D simulation and (b) the 3D simulation. The snapshots are taken at $t=500$ fs and $t = 255$ fs respectively, corresponding to the first peak of electron energy spectra in \cref{fig:Electron_spectrum_2D3D}. The bar width in (a) and (b) is set as 75 MeV.}
\end{figure}

\begin{figure*}[!htb]
    \includegraphics[width=0.99\textwidth,trim={0cm 0cm 0cm 0cm},clip]{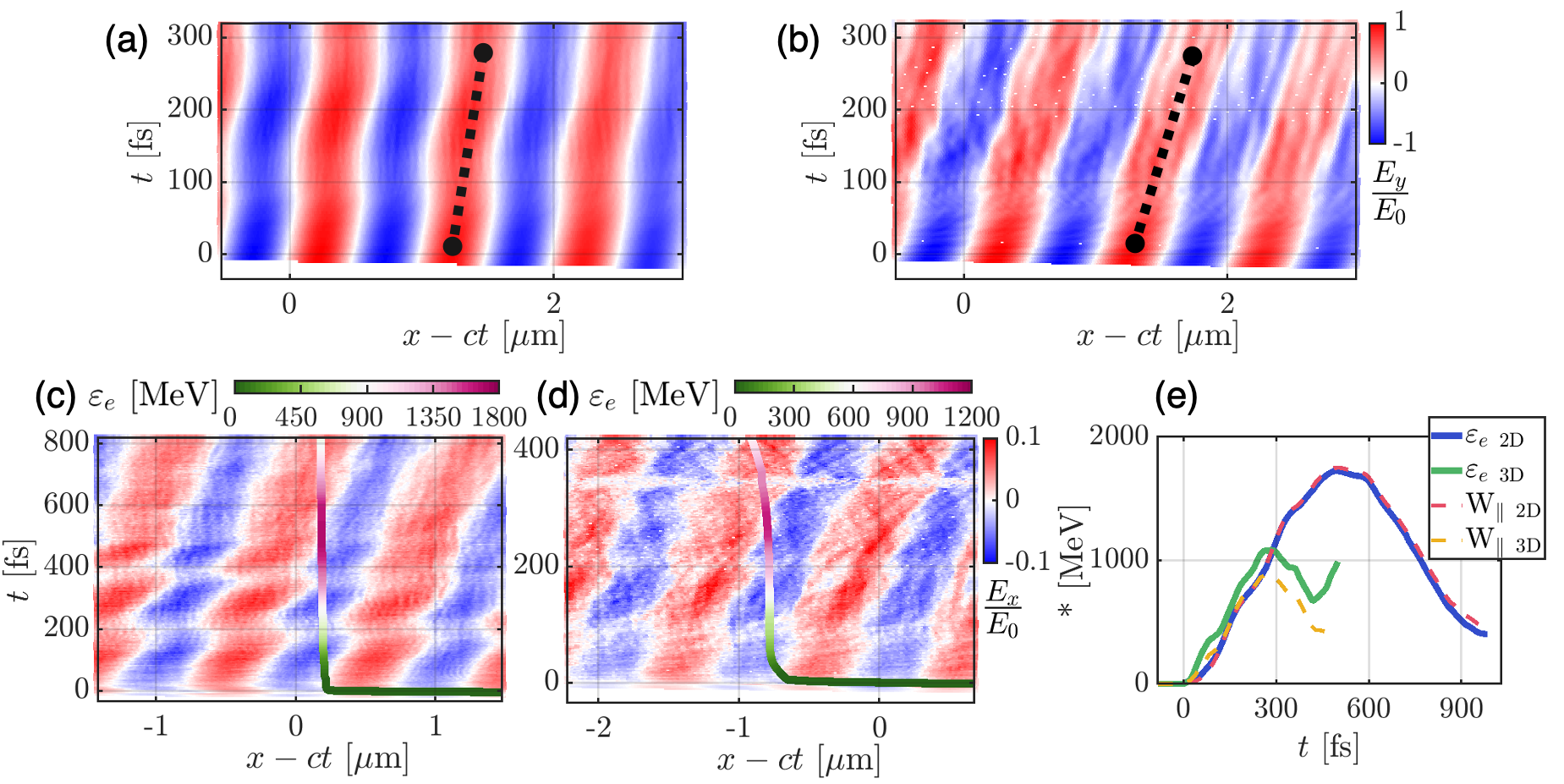}
    \caption{\label{fig:phase_v_traj} (a) Temporal profiles of $E_y$ fields on the central axis in (a) the 2D simulation (recorded at $y=0$) (b) the 3D simulation (recorded at $y=z=0$). The $E_y$ fields are plotted in a moving window which moves with the speed of light. Here $E_0$ with a value of $3.2\times10^{14} \ \rm{V/m}$ is the peak amplitude of the electric field in the incoming laser pulse. The black dashed lines show the segments used to determine $v_{ph}$ in each run. The trajectory of a typical high-energy electron is plotted together with the exact $E_x$ fields sampled by that electron in (c) the 2D simulation (d) the 3D simulation. (e) The comparison of relativistic electron energy and the longitudinal work between the electrons tracked in (c) and (d).}
\end{figure*}

At ultra-high laser intensities, the dephasing rate $\delta u$ is approximately $(v_{ph}-c)/c$, since for a relativistic electron co-propagating with laser, $c-v_x \ll v_{ph}-c$. \Cref{fig:phase_v_traj}(a-b) describes the temporal profiles of the transverse electric fields $E_y$ recorded in a moving window. Note that $E_y$ and $E_x$ share the same wave mode and phase velocity. By tracking a fixed field segment, we find that $v_{ph} \approx 1.0031 \ c$ in the 2D simulation, and $v_{ph} \approx 1.0063 \ c$ in the 3D simulation. The dephasing rate in 2D is nearly two times lower than that in 3D, explaining the distinction between the electron energy gain seen in the two simulations. The duration for electrons staying in an accelerating phase of electric fields can be estimated by 
\begin{equation}
    \delta t \approx 0.5 \lambda_{L} / (\delta u  \cdot c),
\end{equation}
where $0.5\lambda_L$ is the width of the accelerating phase. We find that $\delta t \approx$ 540 fs and 260 fs in 2D and 3D, matching well with the times found from the simulations, demonstrating the accuracy of the approximation used to calculate the phase velocity. The previously derived \cref{Eq:dephasing} however gives the analytical values of the phase velocity as $v_{ph} = 1.0078 \ c$ in 2D, and $v_{ph} \approx 1.0116 \ c$ in 3D. Despite the difference in numerical value, the same trend of an increase in phase velocity is consistent in both the analytical treatment and the numerical simulations. There are a number of factors that can lead to this discrepancy between the analytical calculations and the numerical simulations, for example: the analytical calculation assumes a perfectly conducting boundary and so requires the transverse fields to be zero at the channel edges\cite{Gong_hollow_channel}; sinusoidal waves are not a perfect match (though close) for the fields observed in the hollow channels. Our additional simulations with different target size $R$ show that as phase velocity varies with $R$, the ratio of $\delta u$ in 2D to $\delta u$ in 3D is preserved.

To further check how the phase velocity influences electron acceleration, we track representative electrons with respect to $E_{x}$ fields sampled by them, as illustrated in \cref{fig:phase_v_traj}(c-d). It is clear that the electrons remain accelerated when they stay in the favourable phase of $E_x$ fields (the negative fields colored in blue). After exiting the accelerating phase, the electron energy declines. As can be seen from the time scale in \cref{fig:phase_v_traj}(c-d) electrons in the 2D simulation are subject to far longer periods of acceleration than those in 3D, due to the lower phase velocity in 2D. The evolution of energy and longitudinal work of the chosen electrons is shown in \cref{fig:phase_v_traj}(e). At early moments (up to $\sim$260 fs), the curves of $W_{\parallel_{\ {\rm{2D}}}}$ and $W_{\parallel_{\ {\rm{3D}}}}$ almost overlap with each other implying that the amplitudes of accelerating fields in 2D and 3D are similar. It is then clear that the major cause for the smaller electron energy observed in the 3D simulation is the early termination of the acceleration process due to the higher dephasing rate. 

\section{Influence of dimesionality on photon emission}

The focus of the discussion in the previous sections was the impact of simulation dimensionality on the laser-driven electron acceleration and generation of energetic electrons with energies of hundreds of MeV to a few GeV. These high-energy electrons are subject to emitting energetic photons (in x-ray and even up to the $\gamma$-ray range) while accelerating in laser and channel fields. In this section, we investigate how the photon emission changes with the dimensionality of simulations. The emitted power $P_{\gamma}$ of synchrotron emission is determined by the electron acceleration in an instantaneous rest frame. This acceleration is proportional to a dimensionless parameter\cite{LLclassicfields.1975} $\eta$,
\begin{equation} \label{Eq:eta}
    \eta \equiv \frac{\gamma_e}{E_S} \sqrt{\bigg(\bf{E} + {\rm{\frac{1}{c}}}[v \times B]\bigg){\rm{^2}} - {\rm{\frac{1}{c^2}}}(E\cdot v){\rm{^2}}},
\end{equation}
where $\bf{E}$ and $\bf{B}$ are the electric and magnetic fields acting on the electron, $\gamma_e$ and $\bf{v}$ are the relativistic factor and the velocity of the electron, and $E_S \approx 1.3 \times 10^{18} \ \rm{V/m}$ is the Schwinger field. The emitted power from an electron scales as $P_{\gamma} \propto \eta^2$. We are interested in photons with energy above 100 keV, a threshold shown to be critical for photon-photon pair production~\cite{PRApplied_power}. 

\begin{figure*}
    \includegraphics[width=0.88\textwidth,trim={0.0cm 0.1cm 0cm 1cm},clip]{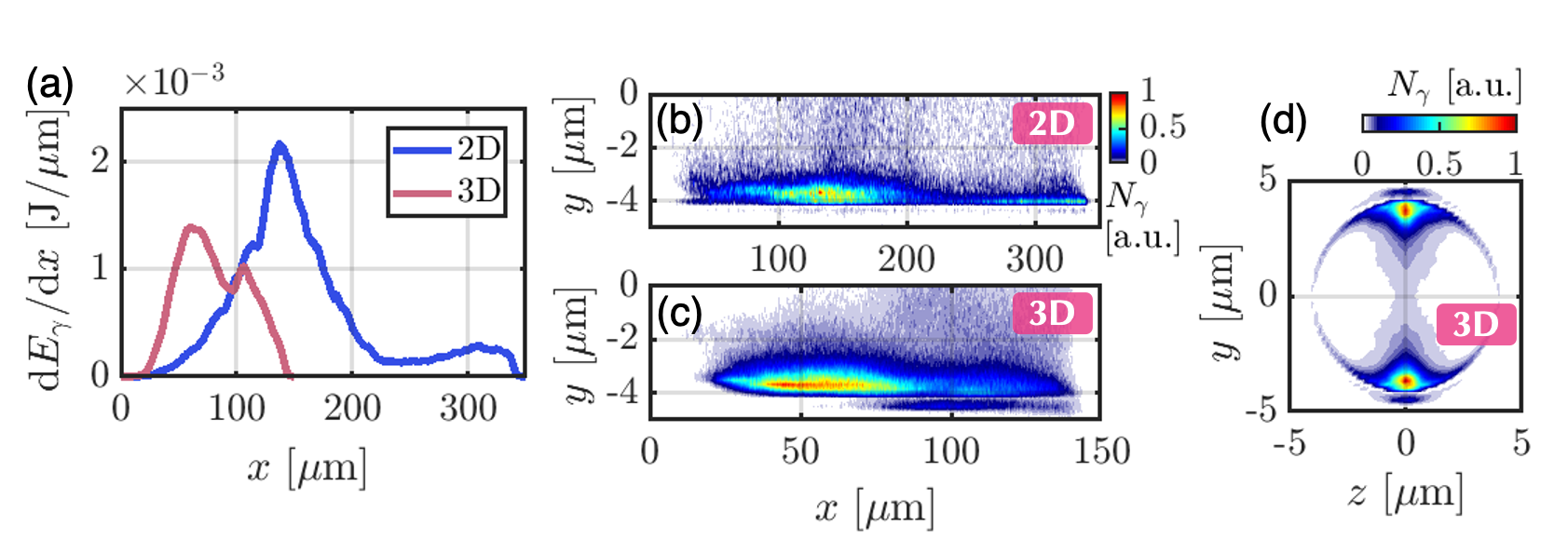}
    \caption{\label{fig:emission_position} Photon distribution over space. (a) Photon energy distribution along $x$ axis. Photon number distribution in ($x$,$y$) space for (b) 2D and (c) 3D. (d) Transverse distribution of photon number for the 3D simulation plotted in ($y$,$z$) space. The considered photons are forward-emitted, accumulated up to 1100 fs and 500 fs for 2D and 3D respectively. The photon energy threshold is 100 keV.}
\end{figure*}

Collecting all the forward-emitted photons over the duration of the simulations, \cref{fig:emission_position} compares the spatial distributions of photons produced in the 2D and 3D simulations. Noting that in the simulations performed here, once a photon is generated, the emission location is marked as the photon location for the duration of the simulation. Corresponding to the electron acceleration, the first peak of emission in 3D takes place at $x \sim 60\ \mu$m, well ahead of that in 2D which is located at $x \sim 140\ \mu$m [see \cref{fig:emission_position}(a)]. Up to the first peaks, 38\% of laser energy in 2D is transferred to particles (photons, electrons and ions) compared to 41\% in 3D. Of the transferred energies 0.47\% is converted into photons ($\varepsilon_{\gamma} > 100 \ \rm{keV}$) in the 2D simulation in contrast to 0.11\% in the 3D simulation. As shown in \cref{Eq:eta}, $\eta$ is directly proportional to electron's relativistic factor, i.e. $\eta \propto \gamma_e$. 
In the previous section it is noted that the field amplitudes acting on the electrons are similar in both 2D and 3D. Thus the primary cause of the lower conversion rate in 3D is due to the lack of electrons at very high energies. \Cref{fig:emission_position}(b-c) illustrates the distribution of photons in ($x$,$y$) space. The common feature of both distributions is that most photons are generated close to the channel boundary, corresponding to the surfing motion of energetic electrons depicted in \cref{fig:Electron_spectrum_2D3D}(d-e). The second peak of emission in both simulations becomes weaker in amplitude and accumulates less photon yield following the first peak. The distribution of the emission in the 3D simulation on the transverse plane ($y$,$z$) is given in \cref{fig:emission_position}(d). It is clear that the emission is concentrated around $z=0 \ \mu\rm{m}$ plane with two populated lobes formed near the channel boundaries, meaning that the emission pattern on the transverse plane is correlated with the direction of laser polarization. 

\begin{figure*}
    \includegraphics[width=1.0\textwidth,trim={0.0cm 0.1cm 0cm 1.02cm},clip]{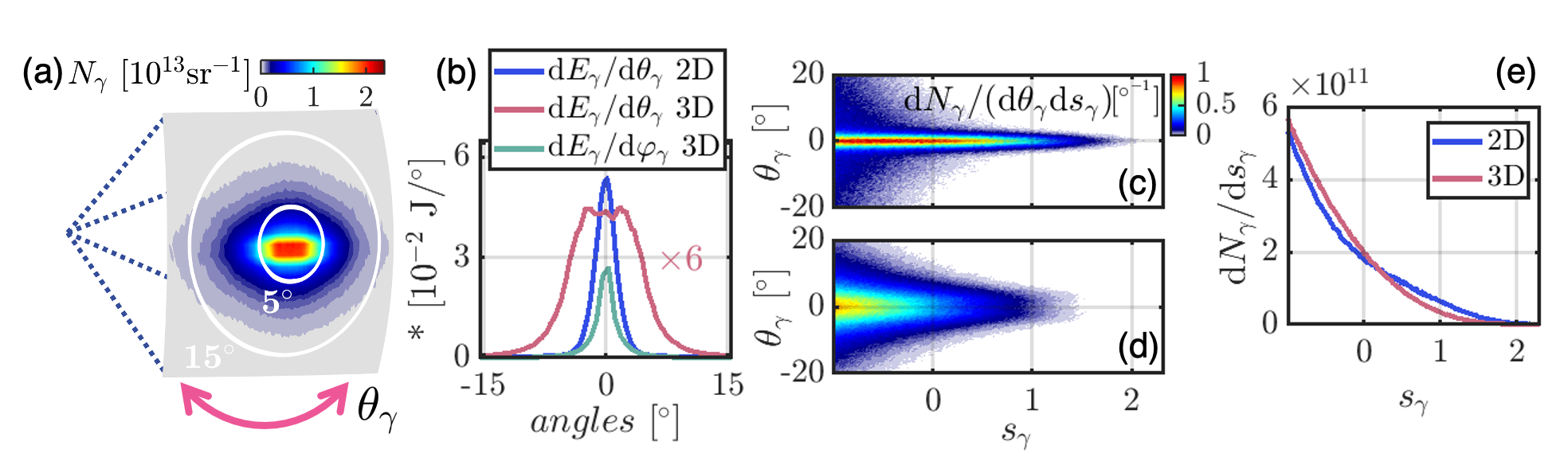}
    \caption{\label{fig:emission_angular} (a) Photon count per solid angle for the 3D simulation. The solid circles mark polar angles of $5^{\circ}$ and $15^{\circ}$. (b) Photon energy distribution over the azimuthal $\theta_{\gamma}$ and polar angles $\varphi_{\gamma}$ for both 2D and 3D, where $\theta_{\gamma} = {\rm{arctan}}(p_y/p_x)$ and $\varphi_{\gamma} = {\rm{arctan}}(pz/\sqrt{p_x^2+p_y^2})$. To make the curve for ${\rm{d}}E_{\gamma}/{\rm{d}}\theta_{\gamma}$ of the 3D simulation visible, its y axis is multiplied by a factor of 6. The spectral-angular distribution of the generated $\gamma$-ray pulse in the (c) 2D and (d) 3D simulations. Here $s_{\gamma} \equiv \rm{log_{10}}(\varepsilon_{\gamma}/MeV)$. (e) Photon energy spectra. The considered photons are forward-emitted, accumulated upto 1100 fs and 500 fs for 2D and 3D respectively. The photon energy threshold is 100 keV.}
\end{figure*}

\Cref{fig:emission_angular} compares the angular distribution and energy spectrum of 2D and 3D simulations. The photon emission in 3D is projected onto a sphere, as illustrated \cref{fig:emission_angular}(a). Though the target is cylindrical, the emission pattern does not preserve the same symmetry. From \cref{fig:emission_angular}(b), the energy distribution along the azimuthal angle ${\rm{d}}E_{\gamma}/{\rm{d}}\theta_{\gamma}$
demonstrates a divergence of $10^{\circ}$ (i.e. the FWHM of the energy distribution curve), which is more than two times wider than that in the polar angle direction. The photon beam in the 2D case is found to be more collimated with a divergence of $6^{\circ}$ in ${\rm{d}}E_{\gamma}/{\rm{d}}\theta_{\gamma}$. Distributing emission over $\theta_{\gamma}$ and photon energy, \cref{fig:emission_angular}(c-d) manifest the photon beam in 2D is better collimated throughout the whole energy spectrum. To further evaluate the quality of the two photon beams, we compare the beam brilliance. The source size of the 3D target is easily decided by its radius while in 2D the source size is set as 4$\times$2 $\mu$m$^2$. It is found that the brilliance of $\gamma$-ray beams is 2.9$\times$$10^{21}$ and 5.8$\times$$10^{20}$ photons/s mm$^2$ mrad$^2$ 0.1\%BW (at 1 MeV) for 2D and 3D respectively. In \cref{fig:emission_angular}(e) starting from 1.5 MeV, the number of $\gamma$-ray photons in 2D surpasses that in 3D, projecting a larger brilliance in 2D for energy level above 1 MeV. 

In this section, we compared the photon yield with regard to the simulation dimensionality. Since photon emission is a direct result of electron acceleration, the emission demonstrates correlated features as observed for electrons in \cref{impact_electron}. The first peak of photon emission in 3D arrives early in space due to the high phase velocity. The lack of high-energy electrons makes the conversion of laser energy to $\gamma$-ray photons less efficient in 3D. Besides the impact on photon number yield, the dimensionality also affects the collimation of the photon beam. With a large divergence angle and a small number of high-energy photons, the 3D $\gamma$-ray beam is less bright than the 2D beam.  

\section{Summary and Conclusions}

We have demonstrated the effects of simulation dimensionality on electron acceleration and $\gamma$-ray
production. There are significant distinctions between the results obtained in 2D and 3D setups from both analytical consideration and numerical calculation. Though in numerical values $v_{ph}$ is close to the speed of light, the dephasing rate ($v_{ph}/c-1$) which determines the energy gain varies considerably with simulation geometry. The higher dephasing rate observable in the 3D setup terminates electron acceleration process early in space and time and leads to a reduction of photon emission when compared to 2D. 

A 2D channel target presents a planar geometry while a 3D target has a cylindrical symmetry. Analytically we have shown that the phase velocity of laser fields propagating inside the targets closely depends on the target geometry; the phase velocity in 2D is smaller than that in 3D. To be more specific, the dephasing rate in a 2D setup is derived to be 1.5 times slower when controlling for the channel radius and laser wavelength. Through numerical simulations, it is found that the absolute majority of work done on energetic electrons comes from the longitudinal electric field, enabling the investigation of laser-driven particle acceleration based only on one single component of the electric fields. By tracking a fixed segment of laser fields the phase velocity in 2D is again shown to be smaller than in 3D, matching the correct trend shown in the analytical derivation. It is clear that in 2D simulations electrons surf for a longer period in an accelerating phase compared to 3D simulations. As a result, electrons in 2D have an elongated acceleration distance and present a more energetic spectrum, leading to an overestimate of maximum electron energy and the number of high-energy electrons when compared to more realistic 3D simulations.

Similar to electron acceleration, photon emission is strongly impacted by simulation dimensionality. Due to a lack of high-energy electrons, in 3D the photon ($\varepsilon_{\gamma} > 100$ keV) conversion rate is more than 4 times smaller and the emission falls behind on the generation of energetic photons ($\varepsilon_{\gamma} > 1.5$ MeV). In 3D, the emission is accumulated close to the $z=0$ plane with a narrow divergence along the polar angle. However the emission along the azimuthal angle is far more diverged in 3D compared to that in a 2D simulation. As a result, the photon beam produced in a 3D setup is found to be less bright. It is also worth noting that the subsequent peaks of photon emission drop sharply in amplitude and duration.

Though 2D numerical simulations are widely applied in the research of laser and micro-channel interactions, one should not ignore the overestimate and inaccuracy of results caused by the low phase velocity in 2D simulations. In particular, when carrying out numerical simulations to predict and optimize the output for laser micro-channel experiments, it matters to accurately know the exact location of peaked electron spectrum and therefore 3D simulations are indispensable. We conclude that the 2D simulations are capable of qualitatively reproducing the features of 3D simulations, but for quantitative evaluations and reliable predictions, 3D modelling is strongly recommended.

\section*{Acknowledgements}
T.W. and A.A. were supported by AFOSR (Grant No. FA9550-17-1-0382). D.B. was supported by NSF (Grant No. PHY 1903098). K.C. was supported by NSF (Grant No. 1821944). Simulations were performed with EPOCH (developed under UK EPSRC Grants No. EP$\/$G054940$\/$1, No. EP$\/$G055165$\/$1, and No. EP$\/$G056803$\/$1) using HPC resources provided by TACC at the University of Texas. This work used XSEDE, supported by NSF grant number ACI-1548562.

\section*{References}



%

\end{document}